\documentstyle[aas2pp4]{article}

\begin{document}

\title{HADRONIC PRODUCTION OF TEV GAMMA RAY FLARES FROM BLAZARS}
\author{Arnon Dar and Ari Laor}
\affil{Department of Physics, Technion-Israel Institute of Technology,
 Haifa 32000, Israel\\
 arnon, laor@physics.technion.ac.il}
\begin{abstract} 
We propose that TeV $\gamma$-ray emission from blazars is produced by
collisions near the line of sight of high energy jet protons with gas
targets (``clouds'') from the broad emission-line region (BLR).  Intense
TeV $\gamma$-ray flares (GRFs) are produced when BLR clouds cross the line
of sight close to the black hole.  The model reproduces the observed
properties of the recently reported very short and intense TeV GRFs from
the blazar Markarian 421. Hadronic production of TeV GRF from blazars
implies that it is accompanied by a simultaneous emission of high energy
neutrinos, and of electrons and positrons with similar intensities, light
curves and energy spectra.  Cooling of these electrons and positrons by
emission of synchrotron radiation and inverse Compton scattering produces
delayed optical, X-ray and $\gamma$-ray flares.
\end{abstract}
\keywords {BL Lacertae objects: general, BL Lacertae objects: individual
(Mrk  421, Mrk  501), gamma rays: theory, radiation mechanisms: non-thermal}

\section{INTRODUCTION}
The Fred Lawrence Whipple observatory has recently reported (Gaidos et al.
1996) the detection of two dramatic outbursts of TeV $\gamma$-rays from
the $\gamma$-ray blazar Markarian 421 (Mrk 421). The first one, on May 7,
1996, had a doubling time of about one hour during which its flux
increased above the quasi-quiescent value by more than a factor of
50. The second outburst, on May 15, 1996, lasted approximately 30 minutes
during which its flux increased by a factor of 20-25. These reports
followed previous reported detections by the Whipple observatory of very
strong bursts of TeV $\gamma$-rays from Mrk 421 which did not seem to be
accompanied by a similar enhancement in GeV $\gamma$-ray emission (Kerrick
et. al 1995; Macomb et al. 1995; Schubnell et al. 1996) and from Markarian
501 (Quinn et al. 1996). Although more than 50 active galactic nuclei
(AGN) have been detected before in GeV $\gamma$-rays by the Energetic
Gamma Ray Telescope Experiment (EGRET) on the Compton Gamma Ray
Observatory (see e.g., von Montigny et al. 1995; Thompson et al. 1995),
all belonging to the blazar type of AGN. Of all the EGRET $\gamma$-ray
blazars, only the nearest, Mrk 421 at redshift z=0.031, has been detected
in TeV $\gamma$-rays (Punch et al. 1992; Lin et al. 1992; Macomb et al.
1995;  Gaidos et al 1996).  Also the blazar Mrk 501 at z=0.034, which is
below the level of detectability by EGRET, has been detected in TeV
$\gamma$-rays (Quinn et al 1996). It has been suggested, that perhaps all
$\gamma$-ray blazars emit TeV $\gamma$-rays, but the opacity of the
intergalactic space to TeV photons due to $e^+e^-$-pair production on the
infrared background photons prevents us from seeing them in TeV photons
(e.g., Stecker et al. 1993). Therefore, it is natural to assume that the
two closest blazars, Mrk 421 and Mrk 501, are not unique and they well
represent TeV $\gamma$-ray blazars. Although Mrk 421 seems to flare in TeV
$\gamma$-rays quite often, most of the EGRET blazars seem to show, within
observational limitations, less variability over very short time scales in
the 30 MeV - 30 GeV energy range (see, however, Mattox et al. 1997). 
 
The observed GeV and TeV $\gamma$-ray emissions from blazars are usually
both interpreted as produced by inverse Compton scattering of highly
relativistic electrons from their jet on soft photons, internal or
external to the jet, (e.g., Maraschi et al. 1992, Bloom and Marscher 1993,
Dermer and Schlickeiser 1993; 1994, Coppi et al. 1993: Sikora et al. 1994;
Blandford and Levinson 1994; Inoue and Takahara 1996).  Although the
radio, X-ray and MeV-GeV $\gamma$-ray emissions are naturally explained by
synchrotron radiation and inverse Compton scattering of high energy
electrons in the jet, there are inherent difficulties in explaining TeV
$\gamma$-ray emission as inverse Compton scattering of soft photons by
highly relativistic electrons or positrons in pure leptonic jets. The main
difficulty is fast cooling of electrons and positrons by inverse Compton
scattering when they are accelerated to TeV energies in the very dense
photon field of an AGN.  Here we would like to propose an alternative
model for TeV emission from blazars based on the assumption that AGN jets
consist of normal hadronic matter (e.g., Mannheim and Biermann 1992).  Tev
$\gamma$-rays are produced efficiently by the interaction of the high
energy protons in the jet with diffuse gas targets of sufficiently large
column density that cross the jet. Such targets may be atmospheres of
bloated stars, stellar winds, or gas clouds in the broad emission-line
region (BLR) around the AGN. We show that the simple properties of
hadronic production of high energy $\gamma$-rays, which are well known
from lab experiments, together with the properties of the BLR of AGNs
which are known from optical, ultraviolet and X-ray studies, explain both
the observed quasi-quiescent emission and the outbursts of TeV
$\gamma$-rays from blazars. We also predict the prompt emissions of TeV
neutrinos, and delayed optical, X-ray and MeV-GeV $\gamma$-ray emissions
that accompany TeV GRFs. 

\section{THE HADRONIC COLLIDER MODEL} 
We assume that $\gamma$-ray blazars are AGNs with highly relativistic jets
of normal hadronic matter that point in the
observer direction. Coulomb coupling of electrons to protons overcomes the
Compton drag in its very dense photon field and makes it possible to
accelerate the jet particles to very large Lorentz factors,
$\Gamma=1/\sqrt{1-\beta^2}\gg 1 $.  For $\Gamma\gg 1$, the kinetic energy
of the jet resides mainly in protons. This energy is converted quite
efficiently into TeV $\gamma$-rays which are beamed towards the observer
by $pp\rightarrow \pi^0X$; $\pi^0\rightarrow 2\gamma$ when ``clouds'' with
high column density from the BLR that surrounds the central region cross
the jet near the line of sight. The quasi-quiescent emission is due to jet
interactions with many, relatively distant, gas ``clouds'' in the BLR.
Strong GRFs are produced when ``clouds'' cross the line of sight at much
closer to the central engine. By ``clouds'' we mean 
diffuse material in the form of atmosphere of bloated stars
(Alexander \& Netzer 1994), stellar
winds or real gas clouds.  Hadronic production of TeV $\gamma$-rays is
accompanied by a simultaneous emission of TeV neutrinos, electrons and
positrons mainly via $pp\rightarrow \pi^{\pm}X$; $\pi^{\pm}\rightarrow
\mu^{\pm}\nu_\mu$;  $\mu^{\pm}\rightarrow e^{\pm}\nu_e\nu_\mu$.  The
subsequent cooling of these electrons and positrons by synchrotron
radiation, inverse Compton scattering and annihilation in flight, produces
optical photons, X-rays and MeV-GeV $\gamma$-rays. 

\section{HADRONIC PRODUCTION OF GAMMA RAYS} 
The cross section for inclusive production of high energy $\gamma$-rays
with a small transverse momentum, $cp_{T}=E_{T}<1~ GeV$ in pp collisions
(e.g., Neuhoffer et al. 1971;  Boggild and Ferbel 1972; Ferbel and Molzon
1974) is well represented by
\begin{equation} 
{E\over \sigma_{in}} {d^3\sigma\over d^2p_{T}
dE_\gamma}\approx (1/2\pi p_{T}) e^{-E_{T}/E_0}~f(x), 
\end{equation} 
where $E\approx m_p\Gamma$ is the incident proton energy, $\sigma_{in} 
\approx 35~mb$ is 
the $pp$ total inelastic cross section at TeV energies, $E_0\approx 0.16~
GeV$ and $f(x)\sim (1-x)^3/\sqrt{x}$ is a function only of the Feynman
variable $x=E_\gamma/E$, and not of the separate values of the energies of
the incident proton and the produced $\gamma$-ray. The exponential
dependence on $E_{T}$ beams the $\gamma$-ray production into $\theta <
E_{T}/E\sim 0.17/\Gamma$ along the incident proton direction. When
integrated over transverse momentum the inclusive cross section becomes
$\sigma_{in}^{-1}d\sigma/ dx\approx f(x).$ If the incident protons have a
power-law energy spectrum, $dF_p/dE\approx AE^{-\alpha}$, then, because of
Feynman scaling, the produced $\gamma$-rays have the same power-law
spectrum:  
\begin{equation} {dF_\gamma\over d E_\gamma}
     \approx N_p \sigma_{in} \int_{E_\gamma}^{\infty} {dF_p\over dE}
      {d\sigma\over dE_\gamma}dE 
      \approx N_p\sigma_{in}IAE_\gamma^{-\alpha},  
\end{equation} 
where $N_p$ is the column density of the
target and $I=\int_0^1x^{\alpha-1}f(x)dx $. 

\section{THE BROAD EMISSION-LINE REGION}
Detailed studies of broad optical and ultraviolet emission-lines, whose
atomic physics is well understood, have been used to obtain detailed
information on the BLR of AGNs. From their line-shapes, relative strengths
and their time-lag response to the variations with time of the central
continuum source, it was concluded that the BLR consists of high column
density broad emission-line clouds (BLCs)
that move with very large random velocities in the BLR:

\noindent    
(i) The size of the BLR has been estimated from reverberation mapping of
both Seyfert 1 galaxies (e.g., Peterson 1993) and quasars (e.g. Maoz,
1997), with typical lag times between 10 days for Seyfert 1 galaxies and
100 days for quasars, respectively. Typically, $R_{BLR}\approx 3\times
10^{16}L_{44}^{1/2}~ cm,$ where $L=L_{44}\times 10^{44}~erg~s^{-1}$ is the
luminosity of the AGN in ionizing radiation. 

\noindent
(ii) The column density and mean density of the clouds were estimated from
the ionizing flux of the central source and the relative line strengths
from the partially ionized clouds. Very high densities and column
densities were inferred. Typical values are, $N_p\sim 10^{23-24}~cm^{-2}$ and
$n_p\sim 10^{10-12}~cm^{-3},  $ respectively. 

\noindent
(iii) For spherical clouds of uniform density $N_c=(4/3)n_c r_c$.
Consequently, the radii of BLCs are typically , $r_c=10^{12}~r_{12}~cm$,
with $r_{12}\sim 0.1-100$. 

\noindent
(iv) The velocity distribution of the BLCs has been estimated
from the profiles of the broad emission lines. Their full widths at half
maximum indicate typical velocities of a few $10^3~ km~s^{-1}$ extending
beyond $10^4~km~s^{-1}$ at the base of the lines. Reverberation mapping
have clearly established that the velocities are not a radial flow (Maoz
1997). They seem to be consistent with the expected velocities of clouds
orbiting massive black holes, $v_c\approx \sqrt {GM/R}\approx 1.15\times
10^9 \sqrt{M_8/R_{16}}~cm~s^{-1}~,$ where $M=M_8\times 10^8M_\odot$ is the
mass of the black hole and $R=R_{16}\times 10^{16}~cm$ is the distance
from the black hole. 
 
\noindent  
(v) The covering factor, i.e., the fraction of the AGN sky covered by
BLCs, was estimated from the ratio of Ly$\alpha$ photons emitted by the
BLCs to the H ionizing photons produced by the central continuum to be,
$C_{BLR}\sim 0.1~.$
 
\noindent
(vi) The total number of BLCs in the BLR was estimated from the sizes of
the BLR and BLCs and the covering factor. Assuming $C_{BLR}\ll 1$, one
finds $N_{BLR}=(4/3) C_{BLR}R^2_{BLR}/r^2_c~.$

\noindent
The UV spectrum of Mrk 421 does not show BLR and $\nu L_\nu(1200~{\rm \AA})\sim
1.3\times 10^{43}h^2~erg~s^{-1}$ (Kinney et al. 1991). 
The BLR may be swamped by beamed
UV power-law emission from the jet. Since the broad lines equivalent
width is $\geq 30$, weaker than in other AGNs, the isotropic
ionizing continuum should be $\leq 10^{42}erg~s^{-1}$ and thus we 
estimate that $R_{BLR}\approx 3\times 10^{15}$ for Mrk 421.

\section{QUASI-QUISCENT EMISSION AND GRFS}   
The hadronic collider model predicts that the TeV $\gamma$-ray emission
from blazars fluctuates with time and shows spectral evolution, even if
the jet properties do not vary with time on short time scales.  
The exact properties of individual flares depend on many unknown
parameters of both the clouds (their geometry, density distribution, speed
and trajectory relative to the jet and line of sight) and the jet (opening
angle $\theta_{jet}$, exact orientation relative to the observer, particle
composition and differential energy spectrum of its high energy particles
as function of distance from the jet axis and along the jet). The general
properties of the quasi-quiescent emission and the flares, however, can be 
estimated using some simplifying assumptions: 
  
Consider a conical jet of particles from a source that is incident on a cloud
at a distance $R$ from the source. Let $b$ and $\theta$ denote their impact 
parameter and angle relative to the jet axis. For the sake of simplicity, 
let us assume that the observer is located at infinity on the
jet axis. Most of the $\gamma$-rays seen by the observer must arrive from 
impact parameters smaller than the critical impact parameter $b_c
\approx RE_0/E_\gamma<R\theta_{jet}$ because of the exponential 
dependence of their production cross section (eq. 1) on $E_T$. 
The number of clouds with $b<b_c$ in the BLR is $N_{BLR}E_0^2/4E_\gamma^2.$
A quasi-quiescent background is formed by jet-cloud interactions 
only if this number is large, i.e., if
$E_\gamma\ll E_{crit}= \sqrt{N_{BLR}}E_0/2\approx 
\sqrt{C_{BLR}} (R_{16}/r_{12})~TeV.$ In that case
the jet  produces a quasi-quiescent $\gamma$-ray flux of
\begin{equation} 
{dF_\gamma\over dE_\gamma}
      \approx C_{BLR} \bar{N}_p\sigma_{in}IAE_\gamma^{-\alpha}.
\end{equation} 
and  the BLR  acts as a target with an effective column
density of
~$C_{BLR}\bar{N}_p$,   
as long as $E_\gamma>E_0/\theta_{jet}$ (below this energy the
produced $\gamma$-rays  are not beamed effectively towards the
observer).  For $E_\gamma > E_{crit}$ the  BLR emission is expected to  
fluctuate considerably.  A flare with a large intensity contrast ratio 
($\equiv$ maximal intensity/quasi-quiescent intensity) is formed when a cloud
crosses the line of sight at a relatively small distance.
If the radius of the cloud is larger than the critical impact
parameter, i.e.,  $r_c>RE_0/E_\gamma$, then when  the cloud
blocks the line of sight, the $\gamma$-ray flux at photon energies 
$E_\gamma>(R/r_c)E_0\sim 1.6R_{16}/r_{12}~ TeV$ flares up with a maximum
intensity, 
\begin{equation} 
{dF_\gamma\over dE_\gamma}
      \approx N_p\sigma_{in}IAE_\gamma^{-\alpha},   
\end{equation} 
where $N_p$ is the average column density of the cloud.  
Thus the maximal intensity contrast  of TeV  GRFs  compared with   
the  quasi-quiescent emission is $N_p/C_{BLR}\bar{N}_p\approx 10-100 $.  
The duration of TeV emission in such flares is of the order of the time it
takes the whole cloud to cross  the line of sight, i.e.,
\begin{equation}
T_{GRF}\sim 2r_c/v_c \sim 1.7\times 10^3 r_{12}R_{16}^{1/2}M_8^{-1/2}~s~.
\end{equation}  
The mean time between such strong flares is 
\begin{equation}
\Delta t\approx \bar{T}_{GRF}R_{BLR}/C_{BLR}b_c\approx 
0.5 L_{44}^{1/2}C_{0.1}^{-1}E_{TeV}^{-1}r_{12}^{-1}T_3~day \end{equation}
where $C_{BLR}=0.1C_{0.1}$ and $\bar{T}_{GRF}=10^3T_3~s.$ 
For $E_\gamma< 1.6R_{16}/r_{12}~TeV$ the maximal GRF intensity is reduced 
by $(r_cE_\gamma/RE_0)^2$ and the duration of the GRF is approximately
the time it takes the cloud to cross the beaming cone:
\begin{equation}
T_{GRF}\sim 2RE_0/v_cE_\gamma\sim 3\times 10^3E_{TeV}^
{-1}R_{16}^{3/2}M_8^{-1/2}~s.  
\end{equation} 
Hence the GRF has the following general behavior when a cloud crosses the
line of sight at a distance $R$:  At energies below $E_\gamma\sim
1.6\times R_{16}/r_{12}~TeV$, the intensity contrast increases with
increasing energy while the duration becomes shorter. Above this energy
both the intensity contrast and the duration become independent of energy.
This behavior results in a spectrum which becomes harder when the
intensity increases and softens when the intensity decreases. The averaged
quasi-quiescent emission spectrum therefore is softer than the spectrum of
strong flares at peak intensity. 

The above predicted properties of the quasi-quiescent emission and the
flaring of blazars in TeV $\gamma$-rays seem to explain quite well those
observed for Mrk 421 and Mrk 501 (Punch et al. 1992;  Lin et al. 1992;
Kerrick et al. 1995; Macomb et al. 1995; Quinn et al. 1996; Gaidos et al
1996). 

\section{NEUTRINOS FROM GAMMA-RAY BLAZARS}
Hadronic production of photons in diffuse targets is also accompanied by
neutrino emission through $pp\rightarrow\pi^{\pm}\rightarrow
\mu^{\pm}\nu_\mu$ ; $\mu^{\pm} \rightarrow e^{\pm}\nu_\mu\nu_e $.  For
a proton power-law spectrum, $dF_p/dE= AE^{-\alpha}$ with a
power index of $\alpha\sim 2$, one finds (e.g., Dar and Shaviv 1996) that the
produced spectra of $\gamma$-rays and $\nu_\mu$'s satisfy
\begin{equation} 
dF_\nu/dE\approx 0.7 dF_\gamma/dE.
\end{equation}
Consequently, we predict that $\gamma$-ray emission from blazars is
accompanied by emission of high energy neutrinos with similar fluxes,
light curves and energy spectra. The number of $\nu_\mu$ events from a GRF
in an underwater/ice high-energy $\nu_\mu$ telescope is $SN_AT_{GRF}\int
R_\mu(d\sigma_{\nu\mu}/dE_\mu)(dF_\nu/ dE)dE_\mu dE$, where $S$ is the
surface area of the telescope, $N_A$ is Avogadro's number,
$\sigma_{\nu\mu}$ is the inclusive cross section for $\nu_\mu p
\rightarrow \mu X$, and $R_\mu$ is the range (in $ gm~cm^{-2}$) of muons
with energy $E_\mu$ in water/ice. For a GRF with $F_\gamma \sim
10^{-9}~cm^{-2}s^{-1}$ above $E_\gamma=1~TeV$ and a power index $\alpha=2$
that lasts 1 day, we predict 3 neutrino events in a $1~km^2$ telescope.
Since the universe is transparent to neutrinos, they can be used to detect
TeV GRFs from distant $\gamma$-ray blazars. If the reported GeV GRF from
the brightest EGRET $\gamma$-ray blazar PKS 1622-297, which had a maximal
flux of $F_\gamma\sim 1.7\times 10^{-5} ~cm^{-2}s^{-1}$ photons above 100 MeV
(Mattox et al 1997), was accompanied by a TeV GRF it could have produced
$\sim 30~\nu_\mu$ events within a day in a $1~km^2$ neutrino telescope. 

\section{X-RAY, MeV and GeV GRFS} 
The production chain $pp\rightarrow
\pi^{\pm}\rightarrow\mu^{\pm}\rightarrow e^{\pm}$ that follows jet-cloud
collisions suddenly enriches the jet with high energy electrons. Due to
Feynman scaling, their differential spectrum is proportional to the
$\gamma$-ray spectrum
\begin{equation} 
dn_e/dE\approx 0.5 dn_\gamma/dE
\end{equation}
and they have the same power-index $\alpha$ as that of the 
incident protons and the produced high energy photons and neutrinos.
Their cooling via synchrotron
emission and inverse Compton scattering produces X-rays, and MeV and GeV
$\gamma$-rays with a differential power-law spectrum
\begin{equation}
dn_\gamma/dE\sim E^{-(\alpha+1)/2}  
\end{equation}  
where $(\alpha+1)/2\approx 1.5-2~.$ Hence, the emission of TeV
$\gamma$-rays is accompanied by delayed emission of optical photons, 
X-rays, and MeV and
GeV $\gamma$-rays.  
The peak emission of synchrotron radiation by electrons with a
Lorentz factor $\Gamma_e$ traversing a perpendicular magnetic field
$H_\perp(Gauss) $ moving with a Doppler factor 
$\delta=(1-\beta
cos\theta)/\Gamma_{H}$ along the jet, occurs at photon energy (Rybicki and 
Lightman 1979) $E_\gamma \sim 5\times
10^{-12} H_\perp\Gamma_e^2\delta ~keV $. The electrons loose $\sim
50\%$ of their initial energy by synchrotron radiation in
\begin{equation}
\tau_c\approx 5\times 10^8 \Gamma_e^{-1} H_\perp^{-2}\delta~s\approx 
1.2\times 10^3H_\perp^{-3/2}E_\gamma^{-1/2}\delta^{-1/2}~s. 
\end{equation} 
Consequently, the time-lag of X-rays is inversely proportional to the
square root of their energy. Similar time-lags for   
MeV-GeV $\gamma$-rays are expected 
if they are produced by Inverse Compton scattering
from the self produced synchrotron photons. The integrated burst energy
over the keV-GeV range is limited by the total electron energy to less than
$\sim 50\%$ of the total energy in the TeV GRF. The spectral evolution of 
the X-ray 
flare (XRF) is a convolution of the spectral evolution of the high energy
electrons and their cooling time. It is hardest around maximum intensity
and softens towards both the beginning and the end of the flare. Because of
electron cooling the spectrum should be harder during rise time than
during decline of the flare. Indeed, all these features have been observed
by ASCA (Takahashi et al. 1996) in the X-ray flare (XRF) that followed the
TeV GRF from Mrk 421 on May 15, 1995. Detailed comparisons will be
presented elsewhere (Dar and Laor 1997). 

\section{SUMMARY AND CONCLUSIONS}
We have proposed an hadronic collider model to explain TeV $\gamma$-ray
emission from blazars. We have used simplifying assumptions to derive from
the model the main properties of quasi-quiescent emission and outbursts of TeV
$\gamma$-rays, neutrinos and X-rays from blazars. The predictions agree
with the Whipple, EGRET and ASCA observations of high energy $\gamma$-ray
emission from Mrk  421 and Mrk  501. This seems to support an hadronic
origin of TeV $\gamma$-rays emission from blazars. Although further
optical, UV and X-ray studies of the broad emission-line region in AGN and
observations of TeV $\gamma$ ray emission from blazars may provide more
evidence for the hadronic nature of AGN jets, a decisive evidence will be
provided by the detection of the predicted TeV neutrino fluxes from gamma
ray Blazars. 

\acknowledgments
Most of this research was done while A.
Dar was visiting the Institute of Astronomy of the University of
Cambridge. He would like to thank the British Royal Society for its kind
support and the members of the Institute of Astronomy for their
hospitality. A. Laor acknowledges support by the ISRAEL SCIENCE
FOUNDATION founded by The Israel Academy of Science and Humanities.

 \end{document}